\documentclass[twocolumn,aps,pra,a4paper,showpacs,preprintnumbers,amsmath,amssymb]{revtex4}

\usepackage[english]{babel}
\usepackage[dvips]{graphicx}
\usepackage{dcolumn}
\usepackage{bm}
\usepackage{amssymb}
\usepackage{mathrsfs}
\begin{document}
\title{Tomographic reconstruction of quantum states in $N$ spatial dimensions}

\author{Anders S. Mouritzen\footnote{Corresponding author}}
 \email{asm@phys.au.dk}
\author{Klaus M\o lmer}
 \email{moelmer@phys.au.dk}
\affiliation{QUANTOP, Danish National Research Foundation Center for
Quantum Optics, Department of Physics and Astronomy, University of
Aarhus, DK-8000 \AA rhus C, Denmark}

\date{\today}

\begin{abstract}
Most quantum tomographic methods can only be used for
one-dimensional problems. We show how to infer the quantum state of
a non-relativistic $N$-dimensional harmonic oscillator system by
simple inverse Radon transforms. The procedure is equally applicable
to finding the joint quantum state of several distinguishable
particles in different harmonic oscillator potentials. A requirement
of the procedure is that the angular frequencies of the $N$ harmonic
potentials are incommensurable. We discuss what kind of information
can be found if the requirement of incommensurability is not
fulfilled and also under what conditions the state can be
reconstructed from finite time measurements. As a further example of
quantum state reconstruction in $N$ dimensions we consider the two
related cases of an $N$-dimensional free particle with periodic
boundary conditions and a particle in an $N$-dimensional box, where
we find a similar condition of incommensurability and finite
recurrence time for the one-dimensional system.
\end{abstract}

\pacs{03.65.Wj}

\keywords{Quantum Tomography, N-dimensional, harmonic oscillator,
box potential, reconstruction}

\maketitle

\section{\label{sec:intro}Introduction}
"What do we know about the state of a physical system given a
certain set of measurements?". While this question permeates all of
physics it is particularly slippery in quantum physics. Here,
measurements disturb a system and potentially alter the outcome of
subsequent measurements. This obstacle is overcome by using a vast
ensemble of uncorrelated and identical quantum systems, where one
measurement is performed on each ensemble member whereafter this
member is discarded. The ensemble must be chosen large enough to
permit measurements of all the quantities of interest and to obtain
statistically significant data about these quantities. Introducing
ensembles this way furthermore cements the role of the quantum state
as the complete statistical information of the quantum system.
Following this ensemble approach, we shall in the present paper
consider the quantum state as being described by the density
operator $\hat \rho$, which can describe both pure and mixed states.
Our goal shall be to find this operator.

In principle, since the quantum state is completely characterized by
its density operator's matrix elements in a complete basis
$\{|\lambda\rangle\}$, one could just measure (the real and
imaginary values of) all these matrix elements $\left\langle
\lambda|\hat \rho|\lambda'\right\rangle$ - i.e. the density matrix.
However, such a general set of observations may be very difficult to
perform experimentally, and instead it will be our aim to find the
quantum state from experimentally realizable measurements.
Specifically, we shall adopt the quantum tomographic approach where
only measurements of the spatial distribution is made at different
points of time
\cite{invradon2}-
\cite{raymerwhitshan}. From these diagonal elements of the density
operator in the position representation, $\textnormal{Pr}(x,t) =
{}_t\hspace{-0.07cm}\left\langle x|\hat\rho|x\right\rangle_t$, and
the known time evolution due to the Hamiltonian we obtain the full
density matrix $\langle x'|\hat \rho|x\rangle$ or, equivalently, the
phase space distribution $W(x,p)$.


Sofar, most methods in quantum state tomography have been concerned
with systems with only one spatial dimension. We shall in this paper
present the extension of two well-known methods of quantum state
tomography to $N$ dimensions. In section \ref{sec:Harmgen} we shall
consider the harmonic oscillator by a treatment similar to that in
\cite{invradon}. This is not a trivial extension as revealed by a
simple consideration of the dimensionality of the sets of
measurements and the quantum state: $\textnormal{Pr}(x,t)$ and
$\left\langle x'|\hat \rho|x\right\rangle$ are both of
dimensionality two in the spatial one-dimensional case, whereas in
the $N$-dimensional case $\textnormal{Pr}(x_1,\ldots,x_N,t)$ is of
dimension $N+1$, but the density matrix $\left\langle
x_1',\ldots,x_N'|\hat \rho|x_1,\ldots,x_N \right\rangle$ is of
dimensionality $2N$.
In section \ref{sec:freepart} we will treat the case of free
particles considered in one dimension in \cite{raymerwhitshan}, but
with periodic boundary conditions and in a box potential. We finally
give a summary of the paper in section \ref{sec:conclusion}.

\section{\label{sec:Harmgen}The Harmonic Oscillator}
It is shown in \cite{glaubercahillordexp} that there is a $1:1$
correspondence between the density operator and the quantum
characteristic function $\widetilde{W}(\xi)$, where $\xi$ is a
complex variable \protect{\footnote{For simplicity, we choose the
ordering parameter $s = 0$ \cite{invradon}. This choice corresponds
to the choice of the Wigner distribution as the complex Fourier
transform of (\ref{eq:1dqcharf}).}}. This means that instead of
directly finding $\hat \rho$ we may just as well find the quantum
characteristic function. This is traditionally the main trick used
in the quantum state reconstruction of the harmonic oscillator.
Before proceeding to the multidimensional case we will briefly
recapitulate this procedure in one dimension. The quantum
characteristic function can be found from $\hat \rho$ by:
\begin{eqnarray} \label{eq:1dqcharf}
\widetilde{W}(\xi) &=& \textnormal{Tr}\left(e^{\xi \hat a^\dagger -
\xi^\star \hat a}\hat \rho\right)
\end{eqnarray}
Under the harmonic oscillator Hamiltonian $\hat H = \hbar
\omega(\hat a^\dagger \hat a + 1/2)$ with angular frequency $\omega$
the ladder operators in the Heisenberg picture evolve according to
$\hat a(t) = \hat a e^{-i\omega t}$, $\hat a^\dagger(t) = \hat
a^\dagger e^{i\omega t}$. Letting $\theta = \omega t$, the position
operator $\hat x$ evolves according to:
\begin{eqnarray} \label{eq:xaft}
\hat x(\theta)      &=& \cos(\theta)\hat x + \sin(\theta)\hat p
\nonumber \\
                    &=& \frac{1}{\sqrt{2}}\left(\hat a^\dagger
e^{i\theta} + \hat a e^{-i\theta}\right).
\end{eqnarray}
Please note that we use dimensionless coordinates $x$ and $p$
\footnote{For the massive harmonic oscillator, this means measuring
$x$ in units of $\sqrt{\hbar/m\omega}$ and $p$ in units of
$\sqrt{\hbar m\omega}$}. We now make a change of variables in
(\ref{eq:1dqcharf}) from the complex number $\xi$ to two real
plane-polar coordinates $(\eta,\theta)$ using $\xi =
\frac{i}{\sqrt{2}}\eta e^{i\theta}$:
\begin{eqnarray} \label{eq:1dqcharfpol}
\tilde{w}(\eta , \theta)    &=& \widetilde{W}\left(\frac{i}{\sqrt{2}} \eta e^{i\theta}\right) \nonumber \\
                            &=& \textnormal{Tr}\left[ e^{i\eta\hat x(\theta)} \hat \rho \right].
\end{eqnarray}
It will prove convenient to choose the variables to be in the
intervals $-\infty < \eta < \infty$ and $0 \leq \theta < \pi$. Note
that the position distribution $\textnormal{Pr}(x,\theta)$ at the
time $\theta/\omega$ is a Fourier transform of the quantum
characteristic function $\tilde{w}(\eta,\theta)$:
\begin{eqnarray}
\textnormal{Pr}(x,\theta)    &=& \textnormal{Tr}\left\{\hat \rho \, \delta \left[\hat x(\theta) - x\right] \right\}\nonumber\\
                &=& \textnormal{Tr}\left\{ \hat \rho \, \frac{1}{2\pi} \int^\infty_{-\infty}\hspace{-0.42cm} d\eta\, e^{i\eta[\hat x(\theta)-
                x]}\right\}\nonumber \\
                &=& \frac{1}{2\pi}\int_{-\infty}^\infty\hspace{-0.42cm} d\eta \, \tilde{w}(\eta,\theta)e^{-i\eta x}.
\end{eqnarray}
By inverting this Fourier transformation, we can find the quantum
characteristic function $\tilde{w}(\eta,\theta)$ from the measured
position distributions:
\begin{eqnarray} \label{eq:1drecon}
\tilde{w}(\eta,\theta) &=& \int_{-\infty}^\infty\hspace{-0.42cm}
dx\,\textnormal{Pr}(x,\theta)\,e^{i\eta x}.
\end{eqnarray}
This is the main equation of quantum tomography in one dimension: We
can find the quantum state (through the quantum characteristic
function) by observing the position distribution for $0\leq \theta <
\pi$ corresponding to one half period of the oscillator.

For completeness we write the result (\ref{eq:1drecon}) in terms of
the Wigner-function; the complex Fourier transform of the
characteristic function \cite{glaubercahillordexp},
\cite{wignerhimself}. The Wigner function is a quasi phase-space
distribution whose marginals along rotated lines are the measured
position distributions:
\begin{eqnarray}
W(x,p)  &=&
\frac{1}{\left(2\pi\right)^2}\int_{-\infty}^\infty\hspace{-0.42cm}d\eta
            \int_0^{\pi}\hspace{-0.25cm}d\theta\,\left|\eta\right|\,\tilde{w}(\eta,\theta)
            \,e^{-i\eta \left[\cos(\theta)x'+\sin(\theta)p\right]}\nonumber\\
        &=& \frac{1}{\left(2\pi\right)^2}\int_{-\infty}^\infty\hspace{-0.42cm} d\eta
            \int_0^{\pi}\hspace{-0.25cm} d\theta \int_{-\infty}^{\infty}\hspace{-0.42cm}dx'
            \,\left|\eta\right|\,\times\nonumber\\
            &&\qquad \textnormal{Pr}\left(x',\theta\right)\,e^{i\eta \left[x-\cos(\theta)x'-\sin(\theta)p\right]}.
\end{eqnarray}

\subsection{The multidimensional oscillator \label{sec:mdharm}}
We will now proceed to show that it is possible to reconstruct the
joint quantum state of a multidimensional harmonic oscillator under
certain conditions.

The Hamiltonian is now $\hat H = \sum_{j = 1}^N \hbar\omega_j(\hat
a^\dagger_j\hat a_j+1/2)$ and we measure the set of $N$ mutually
commuting position operators $\hat x_j$, $j= 1\ldots N$. For
notational simplicity, we arrange these operators in a vector
$\hat{\bm{x}} = (\hat x_1,\ldots,\hat x_N)$. The $N$-dimensional
quantum characteristic function is now $\widetilde{W}(\bm{\xi})$,
where $\bm \xi$ is a vector of $N$ complex variables $\xi_j$. As
before, we let $\xi_j = \frac{i}{\sqrt{2}}{\eta_j} e^{i \theta_j}$,
with $-\infty < \eta_j < \infty$ and $0\le\theta_j < \pi$. This
yields the $N$-dimensional equivalent of (\ref{eq:1drecon}):
\begin{eqnarray} \label{eq:mdrecon}
\tilde{w}(\bm \eta,\bm \theta)&=&\widetilde{W}(\bm \xi)\nonumber\\
                &=& \textnormal{Tr}\left\{\exp\left[i \sum_{j=1}^N \eta_j\, \hat{x}_j(\theta_j)\right] \hat \rho \right\}\nonumber\\
                &=& \int_{-\infty}^\infty\hspace{-0.42cm} d^N\hspace{-0.08cm}\bm{x} \,\textnormal{Pr}\left(\bm x, \bm \theta\right)e^{i \bm \eta \cdot \bm
                x}.
\end{eqnarray}
To gain full knowledge of the function $\tilde{w}(\bm \eta,\bm
\theta)$ we must be able to vary the $N$ variables $\theta_j$
independently of each other on the interval $0\le\theta_j<\pi$. This
is naturally not possible in general, since we can only vary the $N$
variables comprising $\bm \theta$ through variation of the one
parameter $t$. A way to clearly see this restriction is by noticing
that the present reconstruction scheme relies on a
Fourier-transformation, which preserves dimensionality. While we
measure the joint spatial distribution in $N$ dimensions for
different $t$, this equals $N + 1$ dimensional measurements, while
the quantum state (e.g. the characteristic function or Wigner
function) is a $2N$-dimensional object.

It is important to realize what kind of limitations are implied by
the inability to vary the $\theta_j$'s independently. It is always
possible to find the quantum states of a single degree of freedom,
corresponding to tracing out all other degrees of freedom. The
limitation comes about when trying to find the joint quantum state
of the $N$-dimensional system, in particular the correlations
between the different degrees of freedom and entanglement. A simple
illustration of this is offered by a two-dimensional harmonic
oscillator with $\omega_1 = \omega_2$. Let us consider measuring the
observable $\left\langle \hat x_1(\omega t) \hat x_2(\omega t)
\right\rangle$. By using (\ref{eq:xaft}) we find:
\begin{eqnarray}
\left\langle \hat x_1(\omega t) \hat x_2(\omega t)\right\rangle &=&
\mbox{{\large$\langle$}} \left[\cos(\omega t)\hat x_1 + \sin(\omega
t)\hat p_1\right]
\times \nonumber\\
&&\left[\cos(\omega t)\hat x_2 + \sin(\omega t)\hat
p_2 \right]\mbox{{\large$\rangle$}} \nonumber \\
&=& \cos^2(\omega t)\left\langle\hat x_1 \hat x_2 \right\rangle +
\sin^2(\omega t)\left\langle\hat p_1 \hat p_2 \right\rangle +
\nonumber\\
&&\cos(\omega t)\sin(\omega t)\left\langle\hat x_1 \hat p_2 + \hat
p_1 \hat x_2 \right\rangle.
\end{eqnarray}
Here one can see that it is impossible by variation of $t$ to find
the moments $\langle \hat x_1 \hat p_2 \rangle$ and $\langle \hat
p_1 \hat x_2\rangle$, even though one can find their sum. This means
that even a simple two-dimensional Gaussian state cannot be
reconstructed if $\omega_1 = \omega_2$. In section
\ref{sec:kommensurabel} we shall give a precise method to identify
which correlations can be found from a certain set of data. Among
other results we shall see that an $N$-dimensional Gaussian state
can be completely reconstructed if no two $\omega_j$'s are equal.

We return now to the problem with (\ref{eq:mdrecon}): The
$\theta_j$'s are all varied through the one parameter $t$. The
obvious solution to this problem is to devise some means to vary the
$\theta_j$'s independently. There are important situations where
this is indeed possible, e.g. the case of several entangled light
fields. In this case a full reconstruction may be done, regardless
of the values of the $\omega_j$'s, as can be seen in
(\ref{eq:mdrecon}). The independent variation of $\theta_j$ can here
be achieved simply by delaying the measurement on the subsystems by
introduction of, for example, a variable delay line.

Another possibility for varying the $\theta_j$'s independently would
be to vary the times for the subsystems independently. For this
purpose one might use a method closely analogous to the so-called
"twin-paradox" from special relativity. For instance, imagine two
spin-$0$ particles in each their one-dimensional harmonic
oscillator. One may then leave the one subsystem undisturbed while
the other is accelerated to a relativistic speed. This second
subsystem is then allowed to fly along for a while, then accelerated
back again towards the first subsystem and ultimately brought to
rest in its original position. The time dilation will hereby delay
the second subsystem compared to the first, effectively giving a
means to independently vary the elapsed time for the two subsystems.
To avoid direct disturbance of the second oscillator due to the
acceleration, one should accelerate the system perpendicular to its
direction of mechanical oscillation. In principle, this method of
exercising control over the time of subsystems can also be used for
more than two subsystems, but presumably with much increased
practical complication. We note that the idea of relativistic
time-displacement of subsystems has been suggested in
\cite{tvillingkvantprl1990} and recently applied to entanglement
properties with highly non-trivial results \cite{ralphtvilling}.

\subsection{Incommensurable frequencies \label{sec:inkommen}}
\begin{figure}[htbp]
\includegraphics[width=0.5\textwidth]{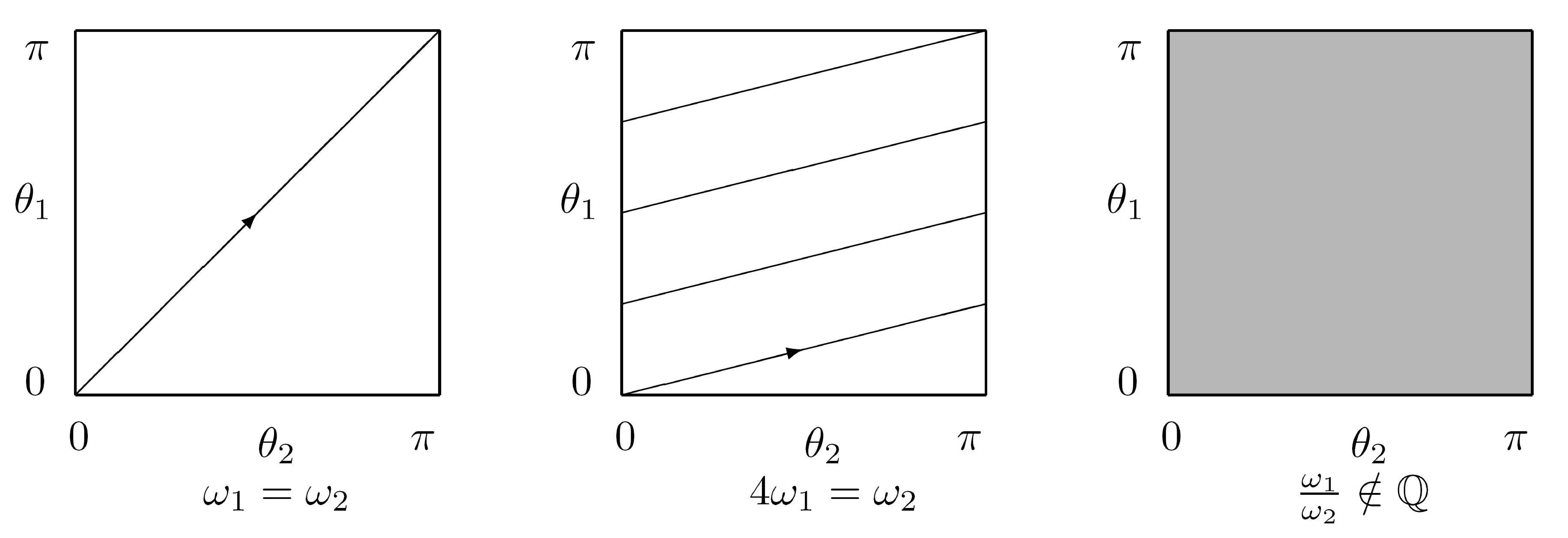}
\caption{\label{fig:kassen} The figure illustrates the variation of
$\theta_j = \omega_j t$ for the case of a two-dimensional harmonic
oscillator. The graph to the left shows what values of $(\theta_1,
\theta_2)$ can be obtained for $\omega_1 = \omega_2$. More
generally, if $\omega_1/\omega_2 = \alpha_1/\alpha_2$ is an
irreducible fraction smaller than $1$, there will be $\alpha_2$
lines in the $(\theta_1,\theta_2)$-plane, as shown in the middle
graph. Moreover, if $\omega_1/\omega_2$ is an irrational number,
almost the whole $(\theta_1,\theta_2)$-plane will be covered as the
measurement time $T' \rightarrow \infty$. The values of
$(\theta_1,\theta_2)$ not covered are all the values where
$\theta_1/\theta_2$ is a rational number. Fortunately, it turns out
we do not need this set of values to exactly reconstruct the quantum
state.}
\end{figure}

Returning to the general problem of full state reconstruction by
(\ref{eq:mdrecon}), we shall discuss under what circumstances this
is indeed possible. By considering all times $t\ge 0$ and choosing
the $\omega_j$'s mutually incommensurable, i.e. their ratios are
irrational numbers, we can find a unique $t$ to reach any $\bm
\theta$ as long as $\theta_j/\theta_k$ with $j \neq k$ is an
irrational number. To see this, remember that $\theta_j =
[\omega_jt]_\pi$, with $[\quad]_\pi$ being the modulus function with
respect to $\pi$. The whole scheme can be pictured as letting $N$
initially coinciding points move around a circle with mutually
incommensurable angular frequencies: If the points coincide at one
angle (which we have chosen to be $\theta = 0$), then no pair will
ever again coincide at this angle. The situation is illustrated for
$N = 2$ in figure \ref{fig:kassen}. A small technical detail in this
respect is that since we have chosen the intervals of $0 \le
\theta_j < \pi$ and $-\infty < \eta_j,x_j < \infty$, then each time
a $\theta_j$ surpasses an integer multiple of $\pi$ we must let
$\eta_j \rightarrow -\eta_j$ in (\ref{eq:mdrecon}).

We have found the function $\tilde{w}(\bm \eta,\bm \theta)$ except
on the values of $\bm \theta$ where two or more
$\theta_j/\theta_k,\, j \neq k$ is a rational number. Fortunately,
this non-available set of $\bm \theta$ values has measure zero, and
since $\tilde{w}(\bm \eta, \bm \theta)$ is uniformly continuous (and
thereby non-singular), the inability to find $\tilde{w}(\bm \eta,
\bm \theta)$ on a set of measure zero is of no consequence. The
uniform continuity of $\tilde{w}(\bm \eta,\bm \theta)$ is a
consequence of $\hat \rho$ belonging to the trace class
\cite{glaubercahillordexp}.

The price we pay to gain knowledge of the $N$-dimensional state as
compared to the one-dimensional case is that 
we must measure the joint position distribution of all coordinates
for all times instead of just half of the oscillators' period. In
actual applications, where infinite measurement times are not
available, one would presumably use frequencies of the $N$
oscillators whose ratios are rational numbers and measure for the
recurrence time of the joint system. The frequencies should then be
chosen so that the $\bm \theta$-space is sufficiently closely
sampled for a reliable reconstruction. The exact amount of
information obtained in such an experiment will be quantified in
subsection \ref{sec:kommensurabel}.

It should also be noticed that this procedure is equally applicable
to a quantum system comprised of several distinguishable
non-interacting particles in separate harmonic potentials.

For completeness we give the formula for reconstruction of the
$N$-dimensional Wigner function:
\begin{eqnarray}
W(\bm x ,\bm p) =
\lim_{T'\rightarrow\infty}\frac{1}{2^N\left(2\pi\right)^{N}T'}
\int_0^{T'}\hspace{-0.38cm} d\bm t\int_{-\infty}^\infty
\hspace{-0.42cm}d^N\bm\eta \int_{-\infty}^{\infty}\hspace{-0.42cm}d^N\bm x'\times\qquad\qquad&&\nonumber\\
             \left|\bm\eta\right|(-1)^{\sum_{j=1}^NF_j(t)}\textnormal{Pr}\left(\bm x',[\bm\omega
             t]_\pi\right)\times \qquad\qquad\qquad\nonumber\\
 \exp{\left\{i\sum_{j=1}^{N}(-1)^{F_j(t)}\eta_j \left[x_j-x'_j\cos([\omega_jt]_\pi)-p_j\sin([\omega_jt]_\pi)\right]\right\}},&&\nonumber
\end{eqnarray}
where $F_j(t) = \textnormal{Floor}(\omega_jt/\pi)$, and the
$\textnormal{Floor}$-function rounds downwards to the nearest
integer.

\subsection{An example of realizing incommensurable frequencies}
In the above discussion, it was demonstrated that the quantum state
of an $N$-dimensional harmonic oscillator could be exactly
reconstructed if the $N$ frequencies were incommensurable. It may be
noticed that the reconstruction of the joint quantum state did not
require interactions between the $N$ degrees of freedom. We will now
give a brief example of how it is possible, by introducing
interactions between the oscillators, to reconstruct the full
quantum state when the all frequencies are identical and equal to
$\omega$. We imagine the $N$ oscillators arranged in a line, and
introduce nearest-neighbor interaction terms in the Hamiltonian. We
let $\kappa \le \omega$ be a real coupling constant:
\begin{eqnarray}
\hat H &=& \sum_{j = 1}^{N} \hbar \omega \left(\hat a_j^\dagger \hat
a_j + \frac{1}{2}\right) + \hat H_{int} \\
\hat H_{int} &=& \sum_{j=1}^{N}\hbar\kappa\left(\hat a^\dagger_j
\hat a_{j+1} + \hat a^\dagger_{j+1} \hat a_j\right).
\end{eqnarray}
Arranging now the $N$ annihilation operators in a column vector $\bm
a = (\hat a_1, \hat a_2, \ldots, \hat a_N)^T$, we find from the
Heisenberg equation of motion:
\begin{eqnarray}
\frac{d}{dt} \left( \begin{array}{c}
\hat a_1(t) \\
\hat a_2(t)\\
\vdots\\
\hat a_N(t)
\end{array} \right)
&=& -i \underbrace{ \left( \begin{array}{cccc}
\omega & \kappa & 0 & \ldots\\
\kappa & \omega & \kappa & \ldots \\
0 & \kappa & \omega & \ldots \\
\vdots & \vdots & \vdots  & \ddots \\
\end{array} \right)}_{\mathscr{D}}
\left( \begin{array}{c}
\hat a_1(t) \\
\hat a_2(t)\\
\vdots\\
\hat a_N(t)
\end{array} \right),\nonumber
\end{eqnarray}
where the matrix $\mathscr{D}$ is tri-diagonal and real. Since there
are no terms containing $\hat a_j^\dagger$, we can perform a usual
orthogonal diagonalization of this matrix to yield $N$ new modes,
characterized by new annihilation operators $\hat a_j'$. The
eigenvalues $\omega '_j$ of $\mathscr{D}$ are well-known from e.g.
H\"uckel molecular orbital theory and solid-state physics:
\begin{eqnarray}
\omega_j' &=& \omega + 2\kappa \cos\left(\frac{2j\pi}{N}\right),
\quad k \in \left\{1, 2, \ldots, N\right\}.
\end{eqnarray}
In this way, the new ladder operators have simple time evolutions
$\hat a_j'(t) = \hat a_j'(0)\exp{(-i\omega'_j t)}$, and we can again
find position operators $\hat x_j'(t) = 1/\sqrt{2}[\hat a_j'(t) +
\hat a_j'{}^{\dagger}(t)]$. So if we choose $\kappa$ so that the
$\lambda_j$'s are incommensurable, we can use the reconstruction
method from subsection \ref{sec:inkommen}. Experimentally, one still
has only to measure the $\hat x_j$'s since the $\hat x_j'$'s are
merely linear combinations of these.

\subsection{Commensurable frequencies and partial information \label{sec:kommensurabel}}
We have shown above that a complete tomographic reconstruction of
the state of an $N$-dimensional quantum oscillator system is
possible if all the oscillator angular frequencies are mutually
incommensurable. This naturally leads to the question of which
aspects of the quantum state can, and which cannot, be obtained from
such tomographic measurements if some of the angular frequencies are
commensurable. We will seek to quantify this degree of information
through the moments of the ladder operators $\hat a_j$ and $\hat
a_j^\dagger$. For this to be meaningful, we must assume that these
moments are finite. For convenience we shall be considering the
Weyl-ordered (i.e. symmetrically ordered) products. Letting $\bm r$
and $\bm s$ be $N$-vectors with non-negative integer components,
these moments are generally of the form:
\begin{eqnarray}\label{eq:wmomenter}
S(\bm{r},\bm{s}) &=&\left\langle \prod_{j=1}^N
\frac{r_j!}{s_j!(r_j-s_j)!} \left\{(\hat
a_j)^{s_j} (\hat a^{\dagger}_j)^{r_j-s_j} \right\}_W\right\rangle, \nonumber \\
\textnormal{where} && r_j \in \mathbb{N}_0 \textnormal{ and } s_j
\in \{0,1,\ldots,r_j\}.
\end{eqnarray}
That is, all factors in the product $S(\bm{r},\bm{s})$ is the sum of
all symmetric permutations of a number $s_j$ of the operator $\hat
a_j$ and a number $(r_j-s_j)$ of the operator $\hat a_j^\dagger$.
Here we let $\{\quad \}_W$ stand for the Weyl ordering, which is the
same as the sum of all permutations divided by the number of terms.
For example:
\begin{eqnarray}
\{\hat a \hat a^\dagger \}_W &=& \{\hat a^\dagger \hat a\}_W =
\frac{1}{2}\left(\hat a \hat a^\dagger + \hat a^\dagger \hat
a\right). \nonumber
\end{eqnarray}
A quantum state is completely characterized if all moments of the
form in (\ref{eq:wmomenter}) are specified. 
In this way we are recasting the question of to what extent the
quantum state can be reconstructed into the question of how many of
the moments $S(\bm r, \bm s)$ can be found.

To keep things transparent, we shall initially consider only two
oscillators with commensurable frequencies and only later generalize
to the case of $N$ oscillators. We shall need the recurrence time
for the system, $T = 2\pi/\omega$ so that $\theta_j = \omega_j t =
\alpha_j\omega t = \alpha_j \theta$, making $\alpha_j$ a positive
integer. Since we are measuring the joint $\bm x$-distributions at
different times, it is natural to consider moments of these
distributions $\left\langle \left[\hat x_1(\theta_1)\right]^{r_1}
\left[\hat x_2(\theta_2)\right]^{r_2} \right\rangle$
\footnote{Actually, no further information on the quantum state can
be gained by considering other functions of the position
distribution. This can be seen by expanding the function in its
moments of the position operators, and realizing that the resulting
equation is merely a linear combination of equations obtained in
(\ref{eq:momogexp})}. Recalling (\ref{eq:xaft}):
\begin{eqnarray}\label{eq:momogexp}
2^{\frac{r_1+r_2}{2}}\left\langle \left[\hat
x_1(\theta_1)\right]^{r_1} \left[\hat x_2(\theta_2)\right]^{r_2}
\right\rangle \qquad \qquad \qquad \qquad \qquad&&\nonumber \\
= 2^{\frac{r_1+r_2}{2}}\left\langle \left[\hat x_1(\alpha_1\theta)\right]^{r_1} \left[\hat x_2(\alpha_2\theta)\right]^{r_2} \right\rangle  \qquad \qquad
\qquad \quad && \nonumber \\
= \left\langle \left(\hat a_1 e^{-\alpha_1\theta} + \hat a^\dagger_1
e^{\alpha_1\theta} \right)^{r_1}\left(\hat a_2 e^{-\alpha_2\theta} +
\hat a^\dagger_2 e^{\alpha_2\theta}
\right)^{r_2} \right\rangle &&\nonumber \\
 = \sum_{s_1 = 0}^{r_1}\sum_{s_2 = 0}^{r_2}
 S(r_1,r_2,s_1,s_2)
 e^{i\theta\left[{\alpha_1(r_1-2s_1)+\alpha_2(r_2-2s_2)}\right]}.
\end{eqnarray}
Since the set of functions $\left\{e^{i n \theta}, n \in
\mathbb{N}\right\}$ is linearly independent on the interval
$[0;2\pi[$ we can find all the moments $S(r_1,r_2,s_1,s_2)$ if there
are no two of the exponential functions in the sum
(\ref{eq:momogexp}) that have the same period in $\theta$. Indeed,
to find all moments (and not only the symmetric ones) of order
$(r_1,r_2)$, we must know all symmetric moments of this and lower
order. A precise way to state this is that there must be no
recurrences in the following lists, where we keep the $r_j'$'s fixed
in each list and let the $s'_j$'s assume all possible values:
$\left(\alpha_1\left[r'_1-2s'_1\right]+\alpha_2\left[r'_2-2s'_2\right]\right)_{r_j'}$
with $r'_j \leq r_j$ and $s'_j \leq r'_j$. Finding these moments can
then be done, for instance, by Fourier transformation since the
aforementioned exponential functions are orthogonal.

Furthermore, one should notice that reconstructing the quantum state
through the moments of the ladder operators, one in principle needs
knowledge only of a small but finite interval of the angle $\theta$,
and not the whole interval $[0;\pi[$. A similar result is found in
\cite{Lmjhydro}. The fundamental assumption that allows for
reconstruction from any small finite $\theta$-interval is that of
finiteness of the moments of the ladder operators and position
operators in (\ref{eq:momogexp}). The earlier discussed method of
state reconstruction via Fourier transformation, (\ref{eq:mdrecon}),
does not suffer from this limitation. On the other hand not all
these moments need be finite, only the ones we use in the
reconstruction.

Finally, one should notice that some of the moments can always be
found, regardless of the value of $\alpha_1/\alpha_2$. This
trivially includes the moments $\left\langle \hat x_1^{r_1} \hat
x_2^{r_2} \right\rangle$, since these are directly measured, but
also the moments $\left\langle \hat a_1^{r_1} \hat a_2^{r_2}
\right\rangle$ and their complex conjugates for any $(r_1,r_2)$
since these moments evolve with the unique largest numerical
frequency.

We now proceed to discuss the practical usefulness of this approach
to reconstruction. It is easy to show that it is possible to find
all moments with $r_1 < \alpha_2$ and/or $r_2 < \alpha_1$, so one
may indeed settle for reconstructing moments of only low order, e.g.
$r_1 + r_2 < \textnormal{max}(\alpha_1,\alpha_2)$. The reason for
this is both the difficulty in precisely measuring higher moments of
the joint position distribution and that finding higher moments of
the ladder operators in general requires the ability to measure very
rapid variations in the joint position distribution (see
(\ref{eq:momogexp})). In addition, it is not necessary to measure a
continuum of angles if one is only interested in moments up to a
certain order, but only a number of angles equalling this number,
which makes the procedure practically feasible. In this way one only
has to solve a number of equations with an equal number of unknowns.

Generalizing the above results to arbitrary $N$, we let $T =
2\pi/\omega$ be the recurrence time of the system and $\theta_j =
\alpha_j\theta$. The positive integers $\alpha_j$ are arranged in an
$N$-vector $\bm \alpha$, and we find that to reconstruct the moments
up to $S(\bm r,\bm s)$ one needs to measure $\left\langle
\prod_{j=1}^N \hat x_j^{r_j}(\theta_j) \right\rangle$ and that there
can be no recurring numbers in each of the lists (once more the
$r_j'$'s are fixed in each list while $s_j'$ assumes all possible
values): $\left(\bm \alpha \cdot \left[\bm{r}'-2\bm{s}'\right]
\right)_{\bm r'}$ with $r'_j \leq r_j$ and $s'_j \leq r'_j$. Like
before, we can in particular always find all moments $\left\langle
\prod_{j=1}^N \hat x_j^{r_j} \right\rangle$, since they are directly
measured, and also $\left\langle \prod_{j=1}^N \hat a_j^{r_j}
\right\rangle$ and its complex conjugate for all $\bm r$.

It is amusing to note that if all the moments of the ladder
operators are finite and the angular frequencies are
incommensurable, it is in principle possible to reconstruct the full
quantum state from measurements made in a small but finite time
interval regardless of the dimension $N$ - an impossible task if
just two angular frequencies are commensurable.

One may also remark that an $N$-dimensional Gaussian state can
always be completely reconstructed if no two $\omega_j$'s are equal:
The required measurements are the joint position distribution for
either $4$ points of time or any finite continuous interval of time.

Lastly, we make a short comment on a possible strategy for guessing
the quantum state in the case of commensurable frequencies. Even
though we do not know the moments of the ladder operators
individually, we still find the sum of two or more - this number
determined by the how many occurrences there are of a particular
number in the aforementioned lists. One method of guessing the state
from an incomplete set of data is the Maximum Entropy Principle, due
to Jaynes \cite{jaynesto}, which has been used in several
reconstruction schemes \cite{Buzekjmo}-\cite{vorartbec}. The Maximum
Entropy principle says that in case one has a set of data which
could have come about due to several different quantum states, one
should choose the state with the largest entropy. In our present
scenario, this means that if we only know the sum of, say, $n_{\bm
r,\bm s}$ different symmetric moments, the Maximum Entropy principle
would ascribe equal Lagrange multipliers to each observable in the
Maximum Entropy density operator.


\section{The free particle\label{sec:freepart}}
The tomographic reconstruction of the completely free particle has
been considered in \cite{raymerwhitshan}, but as stated herein,
cannot be used beyond the one-dimensional case. Here, we instead
study the semi-continuous case of the free particle with different
boundary conditions. In the first two subsections
\ref{sec:1dfreeper} and \ref{sec:mdfreeper}, we shall study the free
particle with periodic boundary conditions, also valid for the
planar rotor \footnote{For the rotor, one must substitute $ma^2
\rightarrow I$, where $I$ is the moment of inertia.}. In the last
subsection \ref{sec:partbox} we shall give a brief treatment of the
particle in a box where it will be seen that the different boundary
condition has an important effect on the available information.

\subsection{The one-dimensional case \label{sec:1dfreeper}}
As in the case of the oscillator, we shall first treat the
one-dimensional case and later extend this to $N$ dimensions. In the
present case the reconstruction of the quantum state will be done
through finding the matrix elements of $\hat \rho$ by simple
inversion of Fourier transforms used in \cite{raymerwhitshan}. Let
us consider a particle with mass $m$ on the spatial interval $0 \le
x < L$. The Hamiltonian is $\hat H = \hat p^2/2m$, yielding the
eigenstates $|n\rangle$ with energy $E(n) = \hbar \Omega n^2$. Here
$\Omega = \pi h/mL^2$ and $h$ is Planck's constant. The eigenstate
$|n\rangle$ in the $x$-representation is:
\begin{eqnarray}\label{eq:1dfreeeigen}
\left\langle x|n \right\rangle_{t} &=& \frac{1}{\sqrt{L}}e^{2\pi i n
x/L}e^{-i \Omega n^2 t}.
\end{eqnarray}
We can use this to find the position distribution at any time:
\begin{eqnarray}\label{eq:1dfreepernnm}
\textnormal{Pr}(x,t) &=& _t\left\langle x| \hat \rho |x\right\rangle_t \nonumber \\
        &\hspace{-1.5cm}=&\hspace{-1.2cm}\sum_{n,n' = -\infty}^{\infty}\hspace{-0.3cm}{}_t\hspace{-0.06cm}\left\langle x|n\right\rangle\left\langle
        n|\hat\rho|n'\right\rangle\left\langle n'|x\right\rangle_t \nonumber \\
        &\hspace{-1.5cm}=&\hspace{-1.2cm}\sum_{n,n' = -\infty}^{\infty}\hspace{-0.34cm}\rho(n,n')\frac{1}{L}e^{2\pi
        i(n-n')x/L}e^{-i\Omega(n^2-n'^2)t}.
\end{eqnarray}
It will now be convenient to change variables from $n$ and $n'$ to
$\bar n = n+n'$, $\Delta n = n-n'$. Note that $\bar n$ and $\Delta
n$ are both either even or odd. Changing summation variables in this
manner yields:
\begin{eqnarray}\label{eq:1dfreepxt}
\textnormal{Pr}(x,t) &=& \frac{1}{L}\left(\sum_{\substack{\bar{n}=-\infty\\
\textnormal{even}}}^{\infty}\sum_{\substack{\Delta n=-\infty\\
\textnormal{even}}}^{\infty} +
 \sum_{\substack{\bar{n}=-\infty\\ \textnormal{odd}}}^{\infty}\sum_{\substack{{\Delta
n}=-\infty\\ \textnormal{odd}}}^{\infty}\right)\times\nonumber\\
        && \rho\left(\frac{\bar{n}+\Delta{n}}{2},\frac{\bar{n}-\Delta n}{2}\right)\times\nonumber\\
        &&e^{2\pi i \Delta{n} x/L}e^{-i\Omega \bar{n}\Delta n
        t}.
\end{eqnarray}
Our task is to invert this equation to find the matrix elements of
$\hat \rho$. Fortunately, this is quite easy. Notice that there are
two exponential functions in (\ref{eq:1dfreepxt}) and both can be
used with a Fourier transformation to pick out certain values of
$\bar n$ and $\Delta n$: The first exponential function in $x$
contains only $\bar n$ and once $\bar n$ is fixed, the other
exponential function can be used to select $\Delta n$.

We let $N_T \in \mathbb{N}$ and $2T = 2\pi/\Omega$ be the minimum
required measurement time, which corresponds to twice the time a
classical particle with the lowest non-zero energy ($n = 1$) would
take to traverse the length $L$. Also, we shall choose $\beta \neq
0$ and obtain:
\begin{eqnarray}\label{eq:1dfreerecon}
&&\int_{0}^{L}\hspace{-0.39cm}dx\,e^{-2\pi i \beta
x/L}\frac{1}{2N_TT}\int_{-N_TT}^{N_TT}\hspace{-0.39cm}dt\,
e^{i\Omega\nu\beta t} \textnormal{Pr}(x,t) = \nonumber \\
&&\frac{1}{2N_TT}\int_{-N_TT}^{N_TT}\hspace{-0.39cm}dt\sum_{\substack{\bar{n}=-\infty\\
\textnormal{parity as $\beta$}}}^{\infty}\rho
\left(\frac{\bar{n}+\beta}{2},\frac{\bar{n}-\beta}{2}\right)\,
e^{i\Omega\left(\nu-\bar{n}\right)\beta t} = \nonumber \\
&&\sum_{\substack{\bar{n}=-\infty\\
\textnormal{parity as
$\beta$}}}^{\infty}\rho\left(\frac{\bar{n}+\beta}{2},\frac{\bar{n}-\beta}{2}\right)\,\delta_{\nu,\bar
n} = \nonumber \\
&& \rho\left(\frac{\nu+\beta}{2},\frac{\nu-\beta}{2}\right) \quad
\textnormal{$\beta$ and $\nu$ of same parity}.
\end{eqnarray}
We have found all elements of the density matrix except those for
which $\beta = 0$, i.e. the diagonal in the (momentum)
$n$-representation. If we were to choose $\beta = 0$, we would
always obtain the result of unity in (\ref{eq:1dfreerecon}), as this
is the same as taking the trace of $\hat \rho$ in the $x$-basis.
This inability to find the diagonal was also pointed out for finite
observation times for the completely free particle in
\cite{raymerwhitshan}, and will unfortunately carry over to the
multidimensional case. This limitation arises because all
probability densities of the eigenstates are identical:
$|\left\langle x|n \right\rangle|^2 = 1/L$ whereby, for example, a
thermal state and any pure eigenstate $|n\rangle\langle n|$ have the
same probability distribution at all times. Even though finding the
diagonal of a density matrix is impossible given the rest of the
matrix, one can use the Schwartz inequality to constrain the size of
the diagonal elements through $|\rho(n,n')|^2 \le
\rho(n,n)\rho(n',n')$.

\subsection{$N$-dimensional free particle with periodic boundary conditions \label{sec:mdfreeper}}
We will now move on to the $N$-dimensional case. We shall work in
the $N$-dimensional interval $x_j \in \left[0,L_j\right[$. The
energies of the eigenstates are $E(\bm n) = \hbar \sum_{j =
1}^{N}\Omega_j n_j^2$ and in extension of (\ref{eq:1dfreepxt}) we
introduce the $N$-vectors of intergers $\bar{\bm{n}}$ and
$\bm{\Delta{}n}$. This yields:
\begin{eqnarray}\label{eq:mdfreeperxt}
\textnormal{Pr}(\bm x,t) &=& \Bigg[\prod_{j=1}^N\left(\sum_{\substack{\bar{n}_j\\
\textnormal{even}}}^{}\sum_{\substack{\Delta{}n_j\\
\textnormal{even}}}^{}+
 \sum_{\substack{\bar{n}_j\\ \textnormal{odd}}}^{}\sum_{\substack{{\Delta{}n_j}\\ \textnormal{odd}}}^{}\right)\times \nonumber\\
&&\frac{1}{L_j}e^{2\pi i \Delta n_j x_j/L_j}e^{-i\Omega_j
\bar{n}_j\Delta n_j t}\Bigg]\times\nonumber\\
&&\rho \left(\frac{\bar{\bm{n}}+\bm{\Delta n}}{2},
\frac{\bar{\bm{n}}-\bm{\Delta n}}{2}\right).
\end{eqnarray}
We shall try to invert this equation to find the matrix elements of
$\hat \rho$. Of course, we cannot approach this completely as in the
one-dimensional case and select a particular vector $\bar{\bm{n}}$
through Fourier-transforms, having only the parameter $t$ to vary.
We can, however, expand the time interval of integration to
incorporate all points of time:
\begin{eqnarray}\label{eq:dobbdelta}
\lim_{T''\rightarrow\infty}
\frac{1}{2T''}\int_{-T''}^{T''}\hspace{-0.39cm}dt\,e^{i\sum_{j=1}^{N}\Omega_j\left(\nu_j-\bar
n_j\right)\beta_j t} = && \nonumber \\
\qquad \qquad
\mbox{{\Large$\delta$}}_{\sum_{j=1}^{N}\Omega_j\left(\nu_j-\bar
n_j\right)\beta_j,\,0}&&.
\end{eqnarray}
If the $\Omega_j$'s are mutually incommensurable, the only
possibility for this Kronecker delta-function to give a non-zero
result is for $\nu_j = \bar n_j \forall j$. Remembering that
$\Omega_j = \frac{\pi h}{m_j L_j^2}$, the condition of
incommensurability of the $\Omega_j$'s is equivalent to demanding
the elements of the list:
\[\left(m_j L_j^2\right)\quad\textnormal{be incommensurable}\]
In this way it becomes possible for a single delta-function to
effectively serve as $N$ distinct delta-functions. We shall
henceforth make this requirement of incommensurability of the
$\Omega_j$'s and, as in the one-dimensional case, we only obtain
useful results for $\beta_j \neq 0$. With these limitations we can
exactly reconstruct the rest of the density matrix. Additionally,
$\beta_j$ and $\bar n_j$ have the same parity for all $j$:
\begin{eqnarray}\label{eq:mdreconfreeper}
\rho\left(\frac{\bm \nu+\bm
\beta}{2},\frac{\bm\nu-\bm\beta}{2}\right) &=&
\lim_{T''\rightarrow\infty}
\frac{1}{2T''}\int_{-T''}^{T''}\hspace{-0.39cm}dt\int_{\bm{0}}^{\bm{L}}
\hspace{-0.39cm}d^{N}\bm{x}\times \nonumber \\
        && e^{-2\pi i \sum_{j=1}^{N}\beta_j x_j/L_j}e^{i\sum_{j=1}^{N}\Omega_j\nu_j\beta_j t}\times\nonumber\\
        && \textnormal{Pr}(\bm x,t),
\end{eqnarray}
The special feature of this system, which allows the treatment
above, is that all its eigenenergies are rational numbers (actually
a integers) times some minimum energy. This can in fortunate
circumstances, i.e. if the products of the energy eigenstates in the
position representation are reasonably placid, allow state
reconstruction equations like that in (\ref{eq:1dfreerecon}). These
reconstruction equations will make use of a finite measurement time,
and by using the trick in (\ref{eq:dobbdelta}), can be used in
multiple dimensions. Another example of this kind of system is the
quantum mechanical rotor with fixed angular momentum projection,
which will be the topic of a forthcoming article \cite{rotafos}.

\subsection{Particle in a box \label{sec:partbox}}
In the last two subsections we found ourselves unable to determine
the momentum distribution for the free particle with periodic
boundary conditions. The reason for this was that all the energy
eigenstates of the system had the same spatial distribution. In this
subsection we shall show how imposing other boundary conditions can
completely alter this situation. In particular, we shall study the
particle in a box, where the spatial density is zero at the
boundaries of the box. As usual, we shall first treat the
one-dimensional case and later move on to the $N$-dimensional case.

We choose $x \in [0,L]$ whereby $E(n) = \hbar \Omega' n^2$, now with
$n \in \mathbb{N}$. We use $\Omega' = \Omega/4 = \frac{\pi h}{4 m
L^2}$, yielding the energy eigenstates:
\begin{eqnarray}
\left\langle x|n \right\rangle_t =
\sqrt{\frac{2}{L}}\sin\left(n\frac{\pi x}{L}\right)e^{-i\Omega' t}.
\nonumber
\end{eqnarray}
These eigenstates all have different spatial distributions, and we
already suspect that we shall be able to reconstruct the momentum
distribution. Following section \ref{sec:mdfreeper} we find the
joint position distribution, similar to (\ref{eq:1dfreepernnm}):
\begin{eqnarray}
&&\textnormal{Pr}(x,t) = \, _t\hspace{-0.07cm}\left\langle x| \hat \rho |x\right\rangle_t \nonumber \\
&&        =\sum_{n,n' = 1}^{\infty}
\rho(n,n')\frac{2}{L}\sin\left(n\frac{\pi x}{L}\right)
        \sin\left(n'\frac{\pi x}{L}\right)e^{-i\Omega'(n^2-n'^2)t} \nonumber \\
        && =\sum_{n,n' = 1}^{\infty}\rho(n,n')\frac{1}{L}\bigg\{\cos\left[(n-n')\frac{\pi
        x}{L}\right] - \nonumber \\
        &&\hspace{3.1cm}\cos\left[(n+n')\frac{\pi x}{L}\right]\bigg\}e^{-i\Omega(n^2-n'^2)t}. \nonumber
\end{eqnarray}
Making once more the substitution $\bar n = n+n'$ and $\Delta n =
n-n'$ we arrive at the box equivalent of (\ref{eq:1dfreepxt}):
\begin{eqnarray}\label{eq:1dpartboxpxt}
\textnormal{Pr}(x,t) &=& \frac{1}{L}\left(\sum_{\substack{\bar{n}=0\\
\textnormal{even}}}^{\infty}\sum_{\substack{\Delta n=-\bar n+1\\
\textnormal{even}}}^{\bar n-1} +
 \sum_{\substack{\bar{n}=1\\ \textnormal{odd}}}^{\infty}\sum_{\substack{{\Delta
n}=-\bar n+1\\ \textnormal{odd}}}^{\bar n-1}\right)\times\nonumber\\
        &&\rho\left(\frac{\bar{n}+\Delta{n}}{2},\frac{\bar{n}-\Delta n}{2}\right)\times\nonumber\\
        &&\hspace{-0.4cm}\left[\cos\left(\Delta n\frac{\pi x}{L}\right) - \cos\left(\bar n \frac{\pi
        x}{L}\right)\right]e^{-i\Omega \bar{n}\Delta n t}.
\end{eqnarray}
Recalling that the set of functions $\{\cos(k\pi x/L)\},
\,k\in\mathbb{N},$ is orthogonal on $x \in [0,L]$ we can use the
cosine functions in (\ref{eq:1dpartboxpxt}) and the exponential
function in time to select a certain term in the sum. Selecting $\nu
\in \mathbb{N}$ and $\beta \in \mathbb{Z}$ with $\nu >
\left|\beta\right|$ and letting $T' = 2\pi/\Omega'$ we obtain the
equivalent of (\ref{eq:1dfreerecon}) \footnote{Performing the
necessary sums, it is easiest to let the sums extend over all
positive and negative $\bar n$ and $\Delta n$ (each sum containing
only even or odd indices) and then in the end set all $\rho(n,n') =
0$ if $n \leq 0$ or $n' \leq 0$.}:
\begin{eqnarray}
&&\rho\left(\frac{\nu+\beta}{2},\frac{\nu-\beta}{2}\right) = \nonumber \\
&& \qquad 2\int_0^{L}\hspace{-0.36cm}dx \cos\left(\beta \frac{\pi
x}{L}\right) \times \nonumber\\
&&\frac{1}{2N_{T}T'}\int_{-N_{T}T'}^{N_{T}T'}\hspace{-0.39cm}dt\,e^{i\Omega'\nu\beta
t}\textnormal{Pr}(x,t), \quad \nu > \left|\beta\right|.
\end{eqnarray}
Here it is apparent that we can reconstruct the full density matrix,
including the momentum distribution. Thus, the choice of boundary
conditions has allowed us to overcome the limitation encountered in
the case of periodic boundary conditions.

Moving on to the multidimensional case where $E(\bm n) =
\hbar\sum_{j=1}^{N}\Omega'_j n_j^2$ we find the equivalent of
(\ref{eq:mdfreeperxt}):
\begin{eqnarray}
\textnormal{Pr}(\bm x,t) &=&
\Bigg\{\prod_{j=1}^N\left(\sum_{\substack{\bar{n}_j = 0\\
\textnormal{even}}}^{\infty}\sum_{\substack{\Delta{}n_j =
\\-\bar n_j+1\\ \textnormal{even}}}^{\bar n_j-1}+
 \sum_{\substack{\bar{n}_j=1\\ \textnormal{odd}}}^{\infty}\sum_{\substack{{\Delta{}n_j} =\\ \bar n_j-1\\ \textnormal{odd}}}^{\bar n_j-1}\right)\times \nonumber\\
&&\hspace{-1.5cm}\frac{1}{L_j}\left[\cos\left(\Delta n_j \frac{\pi
x_j}{L_j}\right)-\cos\left(\bar n_j \frac{\pi
x_j}{L_j}\right)\right]e^{-i\Omega_j
\bar{n}_j\Delta n_j t}\Bigg\}\times\nonumber\\
&&\rho \left(\frac{\bar{\bm{n}}+\bm{\Delta n}}{2},
\frac{\bar{\bm{n}}-\bm{\Delta n}}{2}\right).
\end{eqnarray}
Selecting the two $N$-vectors $\bm \nu$ and $\bm \beta$ with $\nu_j
> \left|\beta_j\right|$ and demanding incommensurability of the elements in the list $\left(m_j
L_j^2\right)$, we find the reconstruction formula corresponding to
(\ref{eq:mdreconfreeper}):
\begin{eqnarray}\label{eq:mdreconfreebox}
\rho\left(\frac{\bm \nu+\bm
\beta}{2},\frac{\bm\nu-\bm\beta}{2}\right) &=&
\lim_{T''\rightarrow\infty}
\frac{1}{2T''}\int_{-T''}^{T''}\hspace{-0.39cm}dt\int_{\bm{0}}^{\bm{L}}
\hspace{-0.39cm}d^{N}\bm{x}\times \nonumber \\
        && \hspace{-1cm}\left[\prod_{j=1}^{N}\cos\left(\beta_j\frac{\pi
        x_j}{L_j}\right)\right]e^{i\sum_{j=1}^{N}\Omega'_j\nu_j\beta_j t}\times \nonumber\\
        &&\textnormal{Pr}(\bm x,t), \qquad \nu_j > \left|\beta_j\right|.
\end{eqnarray}
As in the one-dimensional particle in a box, it is in the
multidimensional case possible to completely reconstruct the density
matrix.

\section{Summary\label{sec:conclusion}}
We have shown how to extend two common schemes of quantum state
tomography from one to $N$ dimensions: The harmonic oscillator and
the free particle on a finite interval. In both cases a complete
reconstruction required extension of the time interval of
observation to all times and required incommensurability of the
eigen-energy differences for the different $N$ spatial dimensions.

For the harmonic oscillator, we quantified the information that can
be reconstructed if some of the $N$ frequencies are commensurable.
This was done through reconstructing the moments of the ladder
operators, and also constitutes a reconstruction method for these
moments if they are finite. This partial reconstruction only
required measurements at a finite number of points of time.

For the free particle on a finite interval we showed that all
off-diagonal elements of the density matrix in the
energy-representation can be reconstructed in the case of periodic
boundary conditions, and that a full reconstruction is possible for
the box potential

\begin{acknowledgments}
We wish to thank associate research professor Uffe V. Poulsen for
useful discussions.
\end{acknowledgments}

\end{document}